\documentclass[sn-nature]{sn-jnl}

\usepackage{graphicx}%
\usepackage{multirow}%
\usepackage{amsmath,amssymb,amsfonts}%
\usepackage{amsthm}%
\usepackage{mathrsfs}%
\usepackage[title]{appendix}%
\usepackage{xcolor}%
\usepackage{textcomp}%
\usepackage{manyfoot}%
\usepackage{booktabs}%
\usepackage{algorithm}%
\usepackage{algorithmicx}%
\usepackage{algpseudocode}%
\usepackage{listings}%
\usepackage{cleveref}
\usepackage[most]{tcolorbox}
\usepackage{ragged2e}

\usepackage{markdown}

\tcbset{
  bot_bubble/.style={
    enhanced,
    colback=blue!10,
    colframe=blue!50,
    boxrule=0.5pt,
    width=0.8\linewidth,
    arc=1mm,
    before skip=5mm,
    drop shadow,
    title=PhenoAssistant,

  },
  user_bubble/.style={
    enhanced,
    colback=gray!10,
    colframe=black!50,
    boxrule=0.5pt,
    width=0.8\linewidth,
    arc=1mm,
    before skip=5mm,
    drop shadow,
    title=User
  }
}

\theoremstyle{thmstyleone}%
\theoremstyle{thmstyletwo}%

\theoremstyle{thmstylethree}%

\usepackage{lineno}
\begin{document}

\title[PhenoAssistant]{PhenoAssistant: A Conversational Multi-Agent AI System for Automated Plant Phenotyping}


\author[1]{\fnm{Feng} \sur{Chen}}\email{Feng.Chen@ed.ac.uk}
\author[1]{\fnm{Ilias} \sur{Stogiannidis}}\email{I.Stogiannidis@ed.ac.uk}
\author[1]{\fnm{Andrew} \sur{Wood}}\email{Andrew.Wood@ed.ac.uk}
\author[1]{\fnm{Danilo} \sur{Bueno}}\email{Danilo.Bueno@ed.ac.uk}
\author[2]{\fnm{Dominic} \sur{Williams}}\email{Dominic.Williams@hutton.ac.uk}
\author[2]{\fnm{Fraser} \sur{Macfarlane}}\email{Fraser.Macfarlane@hutton.ac.uk}
\author[3]{\fnm{Bruce} \sur{Grieve}}\email{Bruce.Grieve@manchester.ac.uk}
\author[4]{\fnm{Darren} \sur{Wells}}\email{Darren.Wells@nottingham.ac.uk}
\author[4]{\fnm{Jonathan} \spfx{A.} \sur{Atkinson}}\email{Jonathan.Atkinson@nottingham.ac.uk}
\author[5]{\fnm{Malcolm} \spfx{J.} \sur{Hawkesford}}\email{Malcolm.Hawkesford@rothamsted.ac.uk}
\author[6]{\fnm{Stephen} \spfx{A.} \sur{Rolfe}}\email{S.Rolfe@sheffield.ac.uk}
\author[7]{\fnm{Tracy} \sur{Lawson}}\email{TLawson@essex.ac.uk}
\author[8]{\fnm{Tony} \sur{Pridmore}}\email{Tony.Pridmore@nottingham.ac.uk}
\author*[8]{\fnm{Mario Valerio} \sur{Giuffrida}}\email{Valerio.Giuffrida@nottingham.ac.uk}
\author[1]{\fnm{Sotirios}  \spfx{A.} \sur{Tsaftaris}}\email{S.Tsaftaris@ed.ac.uk}



\affil[1]{\orgdiv{School of Engineering}, \orgname{University of Edinburgh}, \orgaddress{\street{The King's Buildings}, \city{Edinburgh}, \postcode{EH9 3FB}, \state{Scotland}, \country{UK}}}
\affil[2]{\orgname{James Hutton Institute}, \orgaddress{\street{Errol Road}, \city{Dundee}, \postcode{DD2 5DA}, \state{Scotland}, \country{UK}}}
\affil[3]{\orgdiv{Department of Electrical and Electronic Engineering}, \orgname{University of Manchester}, \orgaddress{\street{Oxford Road}, \city{Manchester }, \postcode{M13 9PL}, \state{England}, \country{UK}}}
\affil[4]{\orgdiv{School of Biosciences}, \orgname{University of Nottingham},  \orgaddress{\street{Sutton Bonington Campus}, \city{Loughborough}, \postcode{LE12 5RD}, \state{England}, \country{UK}}}
\affil[5]{\orgname{Rothamsted Research},  \orgaddress{\street{West Common}, \city{Harpenden}, \postcode{AL5 2JQ}, \state{England}, \country{UK}}}
\affil[6]{\orgdiv{School of Biosciences}, \orgname{University of Sheffield}, \orgaddress{\street{Western Bank}, \city{Sheffield}, \postcode{S10 2TN}, \state{England}, \country{UK}}}
\affil[7]{\orgdiv{School of Life Sciences}, \orgname{University of Essex}, \orgaddress{\street{Wivenhoe Park}, \city{Colchester}, \postcode{CO4 3SQ}, \state{England}, \country{UK}}}
\affil*[8]{\orgdiv{School of Computer Science}, \orgname{University of Nottingham}, \orgaddress{\street{Jubilee Campus}, \city{Nottingham}, \postcode{NG8 1BB}, \state{England}, \country{UK}}}




\newenvironment{preservelinebreaks}
 {\begingroup\obeylines\begingroup\obeyspaces}
 {\endgroup\endgroup}

\newenvironment{user}[1][User]
 {\begin{tcolorbox}[
    enhanced,
    breakable,
    colback=black!5,
    colframe=black,
    title=#1,
    fonttitle=\bfseries
 ]
 \begin{preservelinebreaks}\raggedright}
 {\end{preservelinebreaks}\end{tcolorbox}}


\newtcolorbox[auto counter, number within=section]{assistant}[1][PhenoAssistant]{
    enhanced,
    breakable,
    colback=blue!5,
    colframe=blue!70!yellow,
    title=#1,
    fonttitle=\bfseries,
    listing only,
    listing options={
        style=pythonstyle,
        breaklines=true
    }
}

\newenvironment{system}[1][System Prompt]
 {\begin{tcolorbox}[
    enhanced,
    breakable,
    colback=green!5,
    colframe=green!60!blue,
    title=#1,
    fonttitle=\bfseries
 ]\begin{preservelinebreaks}\raggedright}
 {\end{preservelinebreaks}\end{tcolorbox}}

\lstdefinestyle{pythonstyle}{
    language=Python,
    basicstyle=\ttfamily\small,
    keywordstyle=\color{blue},
    stringstyle=\color{red},
    commentstyle=\color{green!70!black},
    numbers=left,
    numberstyle=\tiny\color{gray},
    showstringspaces=false,
    breaklines=true,
    frame=single
}

\newcommand{\mypar}[1]{\noindent\textbf{#1.}}
\newcommand{\mycode}{\url{https://github.com/fengchen025/PhenoAssistant/}}

\abstract{Plant phenotyping increasingly relies on (semi-)automated image-based analysis workflows to improve its accuracy and scalability. However, many existing solutions remain overly complex, difficult to reimplement and maintain, and pose high barriers for users without substantial computational expertise. To address these challenges, we introduce PhenoAssistant: a pioneering AI-driven system that streamlines plant phenotyping via intuitive natural language interaction. PhenoAssistant leverages a large language model to orchestrate a curated toolkit supporting tasks including automated phenotype extraction, data visualisation and automated model training. We validate PhenoAssistant through several representative case studies and a set of evaluation tasks. By significantly lowering technical hurdles, PhenoAssistant underscores the promise of AI-driven methodologies to democratising AI adoption in plant biology.}

\maketitle
\section{Introduction}\label{sec:introduction}
Plant phenotyping aims to quantify functional and structural traits (phenotypes) of crops and plants, which result from complex interactions between genetics and environmental factors~\cite{fiorani2013future}. Phenotype analysis allows breeders and researchers to disentangle genetic effects from environmental adaptation, enabling the development of crops with improved yields and climate resilience -- an urgent need given that the global population is projected to reach 9.7 billion by 2050 \cite{un_population} and extreme weather events become increasingly frequent \cite{lesk2016influence}.

Accurate plant phenotyping often relies on computational workflows that chain multiple software tools for image processing to measure plant traits and subsequently perform data analysis. However, the steep learning curve associated with these workflows such as programming, machine learning, and data science may prevent
plant practitioners from fully leveraging them~\cite{murphy2024deep}. Moreover, existing workflows typically employ fixed pipelines, which are difficult to extend or modify, restricting their applicability to broader tasks and scenarios. Despite calls for more accessible and versatile phenotyping systems~\cite{coppens2017unlocking}, this need remains inadequately addressed.

A promising approach to bridging the expertise gap and enabling more flexible phenotype analysis is by harnessing the power of large language models (LLMs)~\cite{achiam2023gpt, touvron2023llama}. LLMs have emerged as powerful tools capable of solving diverse tasks from natural language instructions. Rather than focusing on developing computational skills, researchers can focus on critical scientific discovery and exploration, by simply prompting an LLM to address the necessary computational and analytical tasks. However, this promise faces the challenge that LLMs alone often lack the specialised domain knowledge and requisite computational workflows to fulfil user-specified tasks~\cite{shen2024tag, yildiz2024large}. An effective strategy is to augment LLMs with instructions and external tools, forming \textit{AI agents} capable of automatically executing tailored computational routines and data analysis pipelines. Moreover, users should be able to oversee and control the agent’s execution process to correct any potential inaccuracies and misunderstanding by the LLMs. Recently, such agents have successfully been applied across different scientific domains, including chemistry \cite{m2024augmenting,boiko2023autonomous}, material sciences \cite{kang2024chatmof,ghafarollahi2024atomagents} and biology \cite{lei2024bioimage,royer2024omega}. While LLMs have also been applied to agriculture and plant sciences (e.g. agricultural Q\&A~\cite{yang2024pllama}, disease and stress phenotyping~\cite{arshad2024ageval, zhao2024implementation}, genome prediction~\cite{benegas2023dna, mendoza2024foundational}), their potential as AI agents for automating complex and data-intensive phenotype analysis remains unexplored. 

In this study, we introduce PhenoAssistant, the first open-source AI-agent system for automating plant phenotype analysis. By integrating a generalist LLM with a specialised toolkit for plant research, PhenoAssistant allows plant scientists to use free-text task descriptions to prompt a wide range of phenotyping-related data analysis pipelines, such as phenotype extraction from images, phenotypic statistics analysis, and data visualisation. The toolkit features cutting-edge deep learning models and LLM agents to support users’ dynamic needs in extracting and analysing plant traits. PhenoAssistant can be extended (e.g. through built-in model training functions) to accommodate customised and emerging needs. To improve reproducibility, it also allows users to save the generated pipelines and re-execute them on similar datasets. Overall, PhenoAssistant serves as a pioneer for leveraging multi-agent systems in plant phenotyping, demonstrating its potential to streamline complex workflows and democratise AI adoption in plant biology, thereby reducing user barriers and advancing scientific discovery.

\section{Design of PhenoAssistant}
\label{sec:design}
\begin{figure}[t]
    \centering
    \includegraphics[width=1\linewidth]{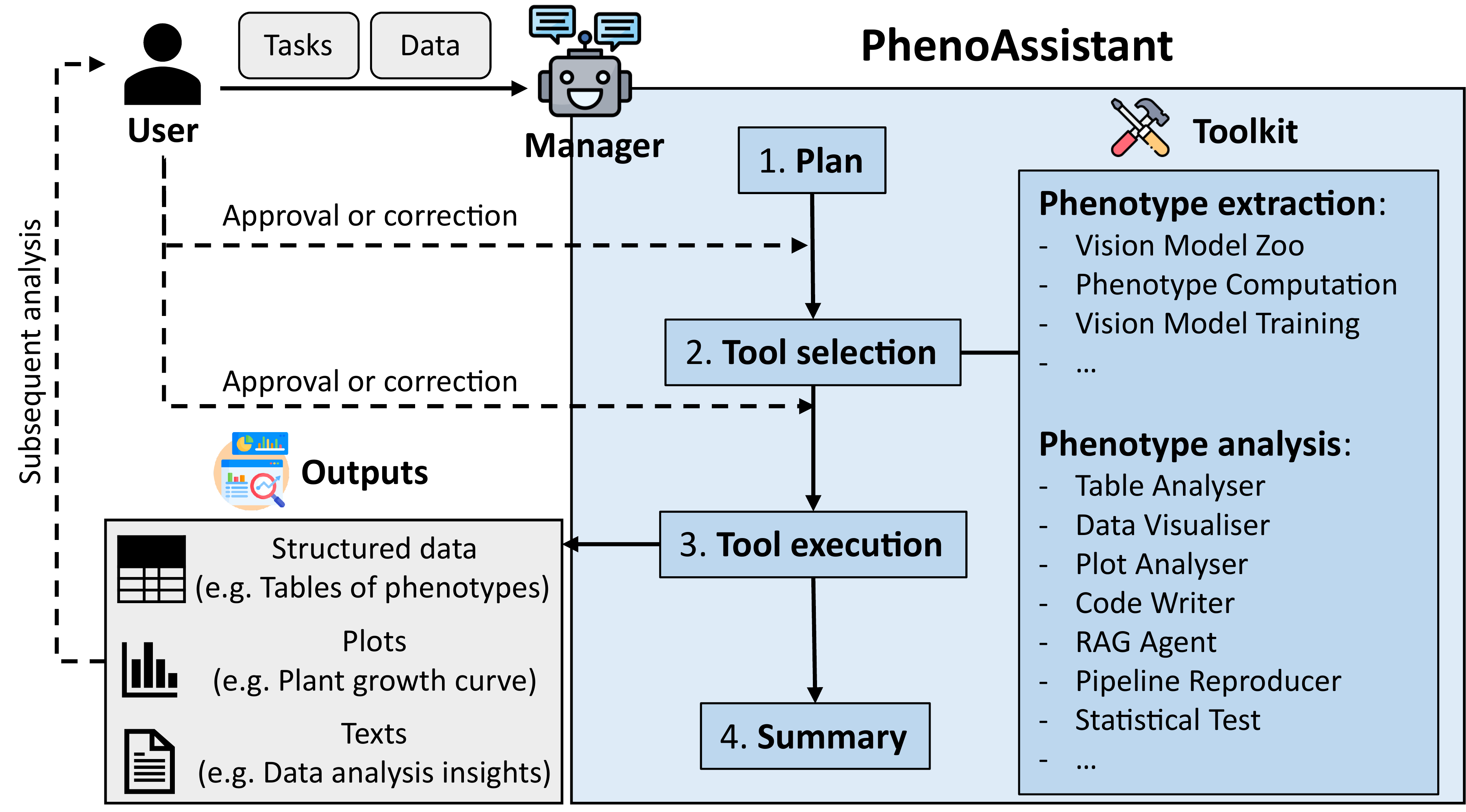}
    \caption{Design of PhenoAssistant. Users provide data and task description to PhenoAssistant. The manager LLM creates a step-by-step plan, selects and executes appropriate tools, then summarises the tool outputs to fulfil the task. Users retain full control to refine intermediate steps as needed. The icon above ``Manager'' is adapted from surang from \url{www.flaticon.com}, and those beside ``Toolkit'' and ``Outputs'' are adapted from Freepik from \url{www.flaticon.com}.}
    \label{fig:overview}
\end{figure}
The goal of PhenoAssistant is to generate and execute suitable pipelines by chaining AI models and other computational tools to address plant phenotype analysis requests from users. To achieve this, it comprises a core LLM (the \textit{manager}) and a specialised \textit{toolkit} designed to enhance its capabilities for image-based plant phenotyping, as illustrated in \Cref{fig:overview}.\footnote{Details of PhenoAssistant's design are presented in Methods (\Cref{sec:methods}).}

The manager coordinates pipeline creation and execution. Upon receiving user-provided task descriptions and accompanying data, it first creates a plan detailing the necessary steps to fulfil the task. It then selects and executes the suitable tools with appropriate parameters, producing various types of outputs that can be saved for further analysis (either by the user or PhenoAssistant). Finally, the manager summarises these outputs, presenting essential information for task completion. Users retain control throughout this process, and can provide textual feedback to refine plans or select alternative tools and parameters, minimising errors from PhenoAssistant. To enhance the capabilities of PhenoAssistant beyond a generalist LLM, we augment it with a toolkit featuring phenotype extraction and analysis.

While recent LLMs often include vision capabilities \cite{achiam2023gpt, anil2023gemini, liu2024visual}, they are not always able to generalise to extracting phenotypes due to the limited representation of plant-specific data in their training datasets.\footnote{We show that ChatGPT cannot correctly extract phenotypes from images at Supplementary \Cref{subsec:gpt}.} For this reason, our toolkit incorporates a \textit{model zoo} of computer vision models trained on plant-specific datasets, with utilities for extracting traits such as plant area and diameter from images. The toolkit also supports automatic model training on new data, enabling users to expand PhenoAssistant's vision capabilities to their unique data and analytical requirements.

Phenotype analysis tools contain various \textit{LLM agents} engineered for different purposes. These include the Table Analyser for extracting information from CSV files, the Data Visualiser for generating plots, and the Plot Analyser for interpreting and analysing generated plots. Additionally, the Code Writer agent supports essential tasks necessary for developing comprehensive end-to-end analysis workflows, such as merging and saving files, as well as executing novel functions not predefined within the toolkit. To enhance PhenoAssistant's expertise in plant-related domains, a retrieval augmented generation (RAG) agent provides access to scientific literature. The Pipeline Reproducer agent enables users to extract and re-execute previously conducted analyses, including all tool calls and generated code, ensuring reproducibility for similar data. Furthermore, tools for statistical test (e.g. ANOVA) are integrated.

\begin{figure}
    \centering
    \includegraphics[width=\linewidth]{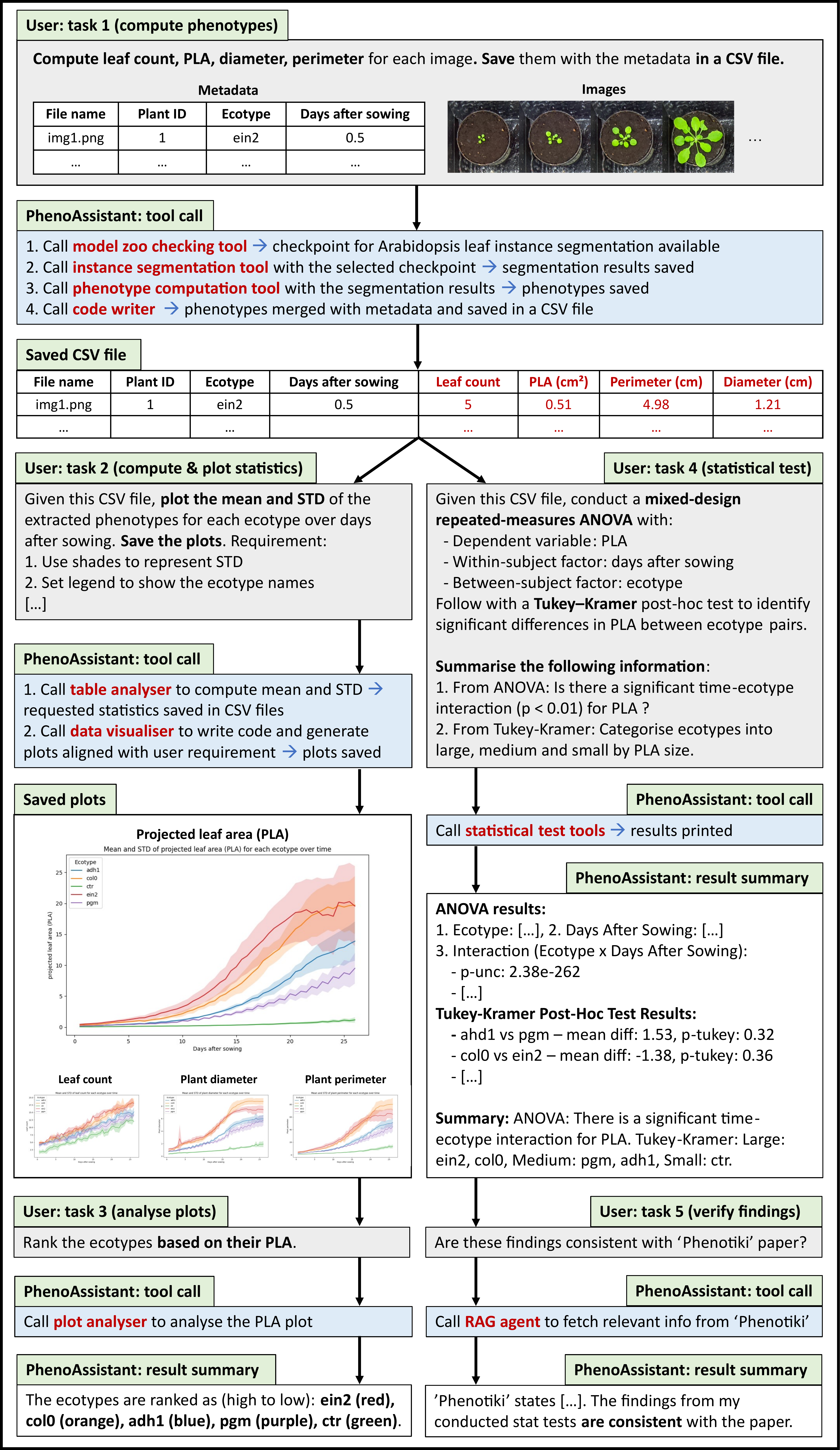}
    \caption{Case Study 1: \textit{A. thaliana} growth pattern analysis. PhenoAssistant automatically completes five tasks: computing phenotypes from images, plotting phenotypic statistics, analysing a generated plot, performing statistical tests for different ecotypes, and comparing findings with literature. Each task is presented as task description (grey), tools used by PhenoAssistant (blue), and results (white).}
    \label{fig:case1}
\end{figure}

\section{Case Studies}\label{sec:results}
Traditionally, executing plant phenotyping tasks requires substantial expertise in computer science, including image processing, machine learning, and coding. With PhenoAssistant, users can complete these tasks effortlessly by interacting through natural language. To demonstrate PhenoAssistant's effectiveness, we present three case studies\footnote{To illustrate how users can interact with PhenoAssistant in practice, we provide the chat logs for \texttt{Case Study 1 Task 1} in Supplementary \Cref{subsec:chatlog}. Complete chat logs for all case studies and tasks are available at \mycode.} showcasing its ability to streamline plant phenotyping analysis.

\mypar{Case Study 1 -- \textit{A. thaliana} growth pattern analysis}
We use PhenoAssistant to replicate part of the Phenotiki \cite{minervini2017phenotiki} analysis pipeline, aiming to visualise and analyse the growth patterns of different \textit{Arabidopsis thaliana} ecotypes. We provide PhenoAssistant with the same dataset as in \cite{minervini2017phenotiki}, which contains 24 plants from 5 different ecotypes: wild type (\textit{Col-0}); \textit{constitutive triple response 1} (\textit{ctr1}); \textit{ethylene insensitive 2} (\textit{ein2.1}); \textit{pgm} (mutation in the plastidic isoform of phosphoglucomutase); and \textit{adh1} (mutation causing defects in alcohol dehydrogenase). Each plant was grown for 26 days with images captured every 12 hours, resulting in a total of 1,248 images. PhenoAssistant does not possess any information about this paper and its pipeline, and our aim is to provide text prompts to obtain comparable results.

As shown in \Cref{fig:case1}, we prompt PhenoAssistant to address five relevant tasks, each including a task description, tools used by PhenoAssistant, and generated results. We begin with the fundamental step of any plant phenotyping task: extracting phenotypes from data: in \texttt{Task 1}, PhenoAssistant is prompted to extract phenotypes including projected leaf area (PLA) and leaf count from images. It identifies the need for a leaf instance segmentation model, and hence selects the one suitable for Arabidopsis from the model zoo, executes it, and computes the requested phenotypes from the segmentation results. The Code Writer is then employed to write and execute Python code to merge the computed phenotypes with metadata and save them into a CSV file. At the end of \texttt{Task 1}, the user can request PhenoAssistant to save the executed pipeline (i.e. from instance segmentation to saving the extracted phenotypes, as demonstrated at Supplementary \Cref{subsec:pipe}), which can later be reapplied to a different dataset.

From \texttt{Tasks 2} to \texttt{5}, we demonstrate PhenoAssistant's capabilities for processing and analysing the extracted phenotypes (i.e. from \texttt{Task 1}), which are essential for identifying plant growth trends and key differences among ecotypes. In \texttt{Task 2}, PhenoAssistant leverages the Table Analyser and Data Visualiser agents to compute and plot statistics based on user-specified plot requirements to visualise the growth patterns of different ecotypes. It can further use the Plot Analyser to interpret data visualisations, such as ranking ecotypes according to their PLA (\texttt{Task 3}). PhenoAssistant can also invoke relevant tools to conduct statistical analyses (e.g. ANOVA and Tukey-Kramer) using user-defined parameters and then summarise the key findings (\texttt{Task 4}). For example, it categorises ecotypes by PLA size as large (\textit{ein2.1}, \textit{Col-0}), medium (\textit{adh1}, \textit{pgm}), and small (\textit{ctr}), using the results from the Tukey–Kramer test. Furthermore, PhenoAssistant can validate these findings against previous literature, by leveraging the RAG agent to embed paper contents and retrieve related knowledge (\texttt{Task 5}).\footnote{For \texttt{Tasks 4} and \texttt{5}, PhenoAssistant performs the same statistical tests and reaches conclusions consistent with the 'Phenotiki' paper \cite{minervini2017phenotiki}.}

\begin{figure}
    \centering
    \includegraphics[width=0.9\linewidth]{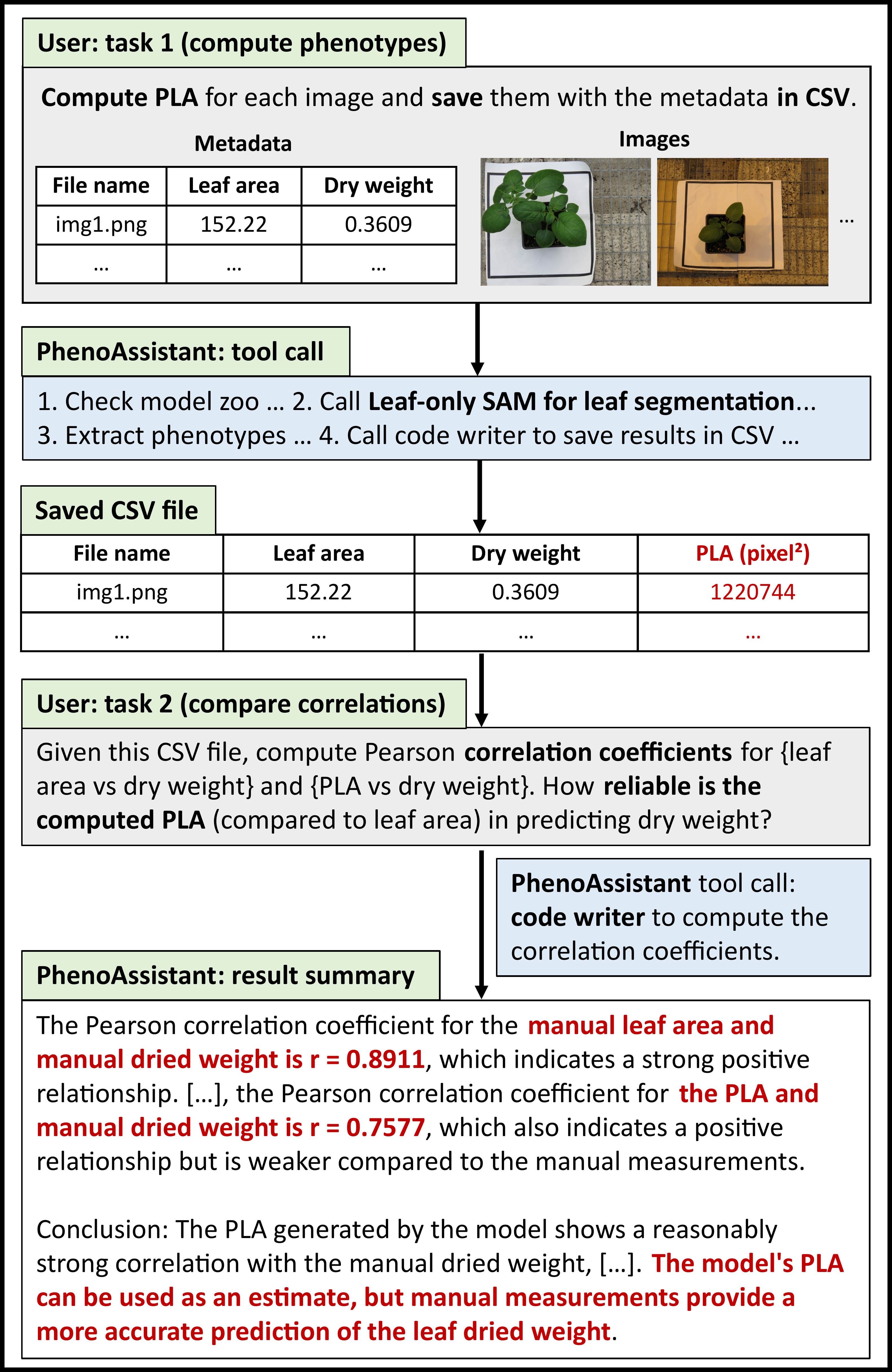}
    \caption{Case Study 2: Potato leaf area and dry weight correlation analysis. In response to the user's requests, PhenoAssistant first extracts phenotypes from the provided data and then compares correlations between different plant-related variables.}
    \label{fig:case2}
\end{figure}

\begin{figure}
    \centering
    \includegraphics[width=0.9\linewidth]{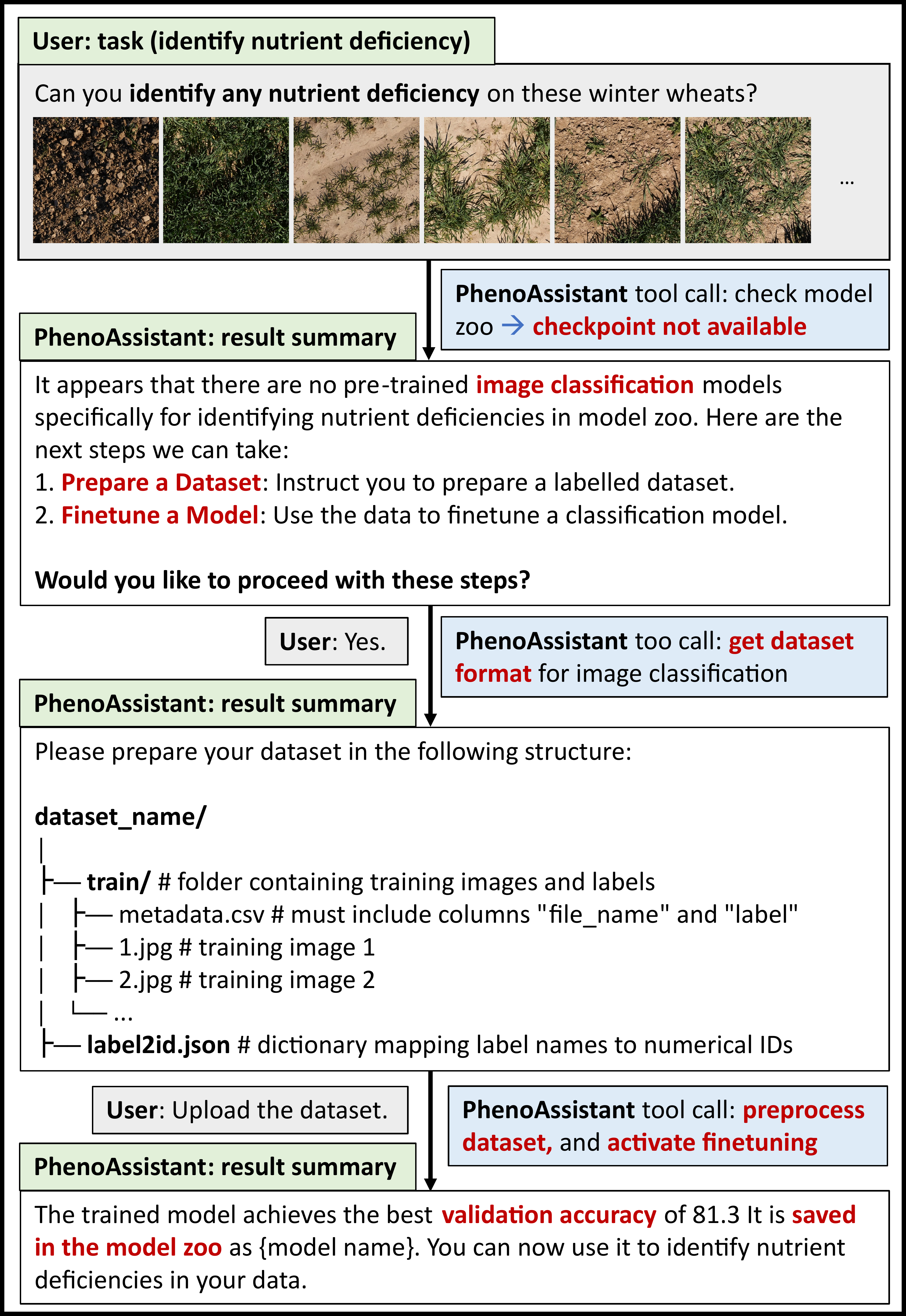}
    \caption{Case Study 3: Automatic model training for nutrient deficiency identification. When no suitable model is available to solve a given task, PhenoAssistant first prompts the user to provide a dataset in the desired format. It then automatically applies data preprocessing, followed by training and evaluating the model. The trained model is saved in the model zoo for future use.}
    \label{fig:case3}
\end{figure}

\mypar{Case Study 2 -- Potato leaf area and dry weight correlation analysis}
PhenoAssistant can handle other data (e.g. non-model plant) and tasks when provided with appropriate tools, such as vision models developed on other datasets. To illustrate this flexibility, we prompt PhenoAssistant to replicate part of the workflow in \cite{williams2024leaf}, assessing the correlation between manually measured leaf area ($A$) and dry weight ($W$), and evaluating if this correlation holds with algorithmically computed $PLA$. This analysis can help plant scientists identify potential discrepancies between manual and automated measurements. For this demonstration, we integrate the Leaf-only SAM model (for potato leaf segmentation) and use a dataset consisting of 32 potato images with $A$ and $W$ annotated; both the model and dataset originate from \cite{williams2024leaf}.

As shown in \Cref{fig:case2}, PhenoAssistant is asked to compute $PLA$ and it follows the similar procedure to the previous case study while using a different segmentation model (\texttt{Task 1}). It is then tasked to measure and analyse the Pearson correlation coefficient between $A$ and $W$, as well as $PLA$ and $W$ (\texttt{Task 2}). As this correlation analysis is not predefined in the toolkit, PhenoAssistant leverages the Code Writer to accomplish it. Finally, PhenoAssistant summarises the results, highlighting the correlation coefficients and providing further explanations and context. The findings indicate that while the $PLA$ derived from Leaf-only SAM is useful, it may introduce errors compared to manually measured leaf area when predicting dry weight.\footnote{PhenoAssistant performs the same correlation analysis and reaches conclusions consistent with \cite{williams2024leaf}.}

\mypar{Case Study 3 -- Model training for nutrient deficiency identification} PhenoAssistant can expand its vision capabilities via the built-in automatic vision model training function, enabling it to tackle previously unseen phenotyping challenges (e.g. experiments with new data or in new environments).

As shown in \Cref{fig:case3}, PhenoAssistant is prompted to identify nutrient deficiency for winter wheats in field, where no capable model is integrated in the toolkit. Recognising the need for image classification, it instructs the user to upload labelled data\footnote{Data used for this demonstration are from \cite{yi2020deep, yi2023non}.} in the required format. After receiving the dataset, it invokes tools to automatically preprocess the data (e.g. splitting into training and evaluation subsets, applying data augmentation), initiate model training, and evaluate the model’s performance. Finally, the evaluation results are presented to the user, and the trained model is added to the model zoo for future use.

\mypar{Summary} We present three case studies illustrating how PhenoAssistant streamlines plant phenotyping with only natural language instructions, allowing plant researchs to concentrate on scientific discovery rather than technical training. These cases span various plant species, environmental conditions, and tasks, including phenotype extraction, pipeline reproduction, statistics computation and visualisation, plot analysis, statistical testing, knowledge retrieval, correlation assessment, and model training.

\section{Evaluation of PhenoAssistant}
Beyond the presented case studies, we evaluate PhenoAssistant’s adaptability and potential for broader applications via a series of tasks. Since PhenoAssistant leverages rapidly evolving AI technologies, this evaluation also offers a method for assessing new AI components (e.g. DeepSeek \cite{bi2024deepseek, guo2025deepseek}) as replacements for PhenoAssistant's original modules. The evaluation covers three aspects\footnote{Prompts used for these evaluation tasks are presented at Supplementary \Cref{subsec:eval_prompts}. Complete chat logs and evaluation results are available at \mycode.}:
\begin{enumerate}
    \item \textbf{Tool Selection (10 Tasks):} We assess whether PhenoAssistant can select appropriate tools with the correct parameters and in the correct order to complete a task. This evaluates PhenoAssistant's abilities of chaining multiple tools for task completion rather than the correctness of task outputs.
    \item \textbf{Vision Model Selection (50 Tasks):} We evaluate whether PhenoAssistant can recommend the appropriate type of computer vision model (instance segmentation, image classification, or image regression) for a given plant phenotyping task. This evaluation is crucial for automatic model training, as users may not have substantial machine learning knowledge to associate plant tasks with computer vision models. Accurate suggestions from PhenoAssistant prevent users from preparing incorrect training data. 
    \item \textbf{Data Analysis (10 Tasks):} For data analysis requests (extracting values from CSV files, computing statistics, and generating plots), we evaluate whether PhenoAssistant can produce the correct outputs. This aims to assess the reliability of the built-in LLM agents (e.g. data visualiser) for data analysis.
\end{enumerate}

During the evaluation, no human feedback (e.g. corrections to tool selection or tool arguments) is provided.

Overall, PhenoAssistant successfully completes almost all tasks. It has a success rate of $70\%$ in tool selection; the failure cases are due to misinterpretation of the tool descriptions and how it should be used, such as using a tool that cannot extract organ length to attempt wheat spike length extraction. It succeed in 98\% vision model selection tasks, with the error arising in an ambiguous case: it recommends using image classification for plant vigor scoring, while regression would be more appropriate for generating continuous scores. All data analysis tasks are completed successfully.

In summary, our evaluation shows that PhenoAssistant, as a multi-agent AI system, exhibits promising adaptability to broader plant phenotyping scenarios and tasks.

\section{Conclusion}\label{sec:conclusion}
We introduce PhenoAssistant, a pioneering AI-powered system designed to streamline plant phenotyping workflows for plant practitioners without requiring computational expertise. By integrating cutting-edge AI technologies, PhenoAssistant transforms simple, interactive conversations into complex image-based plant analyses. Our case studies demonstrate its versatility across a range of phenotyping tasks and plant species, while evaluations highlight its high success rates in tool selection, vision model selection, and data analysis. Overall, PhenoAssistant marks a significant step toward democratising AI adoption in plant phenotyping. It empowers researchers and practitioners to analyse plant data efficiently without requiring deep technical expertise and lays a solid foundation for future AI-driven advancements in plant science.

Despite its success, PhenoAssistant has limitations, primarily due to current LLMs' constraints, which may require complex tasks to be carefully instructed or broken down into multiple steps. This can be improved with more effective prompt engineering or fine-tuning. We also integrate a function to reproduce executed pipelines, reducing the user-LLM interaction for these complex tasks. While our studies demonstrate its support for various species and tasks, its adaptability to rare scenarios requires further investigation. To facilitate ongoing development and validation, PhenoAssistant is made available as open source.

\section{Methods}\label{sec:methods}
PhenoAssistant integrates state-of-the-art AI technologies to allow plant practitioners to perform plant phenotyping through natural language interactions. This section describes the methodological and technical details of PhenoAssistant.

\subsection{Preliminaries: LLMs and Agents}
LLMs are advanced AI models with billions of parameters trained on large text datasets to understand and generate human-like texts. They can assist users in diverse tasks such as writing, coding, and question-answering. Recent advancements, such as GPT-4o \cite{achiam2023gpt}, have also expanded their capabilities to visual understanding and complex reasoning, making them suitable to support more complex tasks.

The versatile abilities of LLMs make them effective as foundations for autonomous systems (i.e. agents). Agents leverage LLMs as the core intelligence for reasoning, decision-making, and performing actions to achieve specified goals. In such systems, LLMs can manage task planning and coordinate interactions with external tools, or directly perform specific functions such as coding and data visualisation. The roles and behaviours of these agents can be crafted by carefully designed system prompts.

\subsection{Details of PhenoAssistant}
PhenoAssistant comprises a \textit{manager} LLM and a specialised \textit{toolkit} designed for plant phenotyping tasks. The manager provides an interactive interface for handling user requests, planning tasks, selecting and executing tools, and summarising results. It is implemented using GPT-4o with a crafted system prompt (Supplementary \Cref{subsec:system_prompts}) that instructs it to ensure desired functionality and behaviour.

The toolkit consists of Python-based functions, each accompanied by a clear function description and detailed explanations of its parameters. These descriptions enable the manager LLM to understand the tools' capabilities and utilise them properly. We introduce the key tools below; other tools can be accessed in our code repository.

\mypar{Vision Model Zoo}
Many plant phenotypes cannot be extracted without the aid of image processing or computer vision techniques. Although recent LLMs have incorporated vision capabilities, they are not yet powerful enough for effective extraction of plant traits. For example, tasks such as delineating individual leaf boundaries (known as instance segmentation in computer vision), which is a preliminary task for computing leaf area and count, remain challenging (Supplementary \Cref{subsec:gpt}).

To overcome this limitation, PhenoAssistant integrates external computer vision models. Each vision model is uniquely identified using the naming convention: \texttt{\{plant-species\}\_\{task\}\_\{(optional) training-dataset\}\_\{model\}\_\{(optional) finetuning-method\}}. These identifiers are maintained within a structured file (e.g. \texttt{model\_zoo.json}), which the manager LLM accesses whenever a vision model is required, ensuring appropriate selection for the given plant task.

For demonstration purposes (\textit{c.f.} \Cref{sec:results}), in \texttt{Case Study 1}, we integrate Mask2Former~\cite{cheng2022masked} for segmenting individual Arabidopsis leaves. Following the similar practice in~\cite{chen2024gmt}, Mask2Former is fine-tuned on subsets A1 and A4 of the publicly available CVPPP LSC dataset~\cite{minervini2016finely}. In \texttt{Case Study 2}, we integrate Leaf-only SAM~\cite{williams2024leaf} for potato leaf instance segmentation.

\mypar{Automatic Vision Model Training}
Given the diverse conditions encountered in plant phenotyping, such as varying illumination, indoor/outdoor environments, species diversity, and diverse phenotyping needs, it is practically impossible to pre-integrate every required vision model into PhenoAssistant. Currently, there is also no universal vision model capable of solving all phenotyping tasks. Furthermore, new tasks and customised user requirements continuously emerge, highlighting the need for a convenient mechanism to expand PhenoAssistant’s vision model zoo.

To satisfy this need, PhenoAssistant incorporates an automated model training pipeline for common phenotyping tasks (e.g. image classification). When users identify the need for a new vision model, or when PhenoAssistant determines that its current model zoo cannot adequately solve a provided task, it prompts the user to upload a dataset formatted according to predefined specifications. PhenoAssistant calls tools to automatically split the uploaded dataset into training and validation sets and initiate model training. Upon completion, the newly trained model is automatically added to the vision model zoo using the naming convention described earlier and becomes available for inference.

The core idea behind this automated training pipeline is to fine-tune pre-trained vision models using plant-specific datasets~\cite{chen2023adapting}. These models are initially trained on large-scale general data and can be fine-tuned to specific plant analysis tasks with limited additional data, thereby minimising development and deployment efforts. Specifically, for image classification (as shown in \Cref{sec:results} \texttt{Case Study 3}), we employ the DINOv2-base model~\cite{oquab2023dinov2} as the pre-trained model.

PhenoAssistant supports two fine-tuning strategies: low-rank adaptation (LoRA)~\cite{hu2022lora} and full fine-tuning, to accommodate users with varying levels of computational resources. LoRA updates only a small subset of parameters by inserting low-rank trainable matrices into existing model layers, substantially reducing computational and memory requirements during training. This approach enables rapid adaptation at the cost of a modest performance trade-off. In contrast, full fine-tuning updates all model parameters, potentially achieving higher accuracy while at increased computational and memory demands.

\mypar{LLM Agents for Phenotype Analysis}
Beyond phenotype extraction, PhenoAssistant supports diverse data analysis tasks such as statistics computations, data visualisation, and interpretative plot analysis. These analyses often involve complex and customised requirements that cannot always follow pre-defined logic. For example, users may request specific visualisation styles (e.g. plotting plant growth curves for different ecotypes using specific colour), or novel analyses prompted spontaneously during exploration.

To support these dynamic and evolving analytical needs, we repurpose GPT-4o into different roles using different system prompts (Supplementary \Cref{subsec:system_prompts}). Each role is wrapped in a Python function to be called by the manager LLM.
\begin{enumerate}
    \item \textbf{Code Writer:} Generates and executes Python code for performing tasks that are not previously defined in the toolkit.
    \item \textbf{Data Visualiser:} Generates and executes Python code for producing plots with user specification. Guidance for creating clear and well-designed plots (e.g. including legends and beautifying layout) is provided within its system prompt.
    \item \textbf{Table Analyser:} Queries values and computes statistics from CSV files using Pandas AI. Pandas AI is a Python library that integrates LLMs directly with the Pandas data analysis library, allowing natural language queries for data analysis.
    \item \textbf{Pipeline Reproducer:} Enhances result reproducibility by extracting executed function calls and code from the chat history, organising them into a new Python function that can be reused in the future. 
    \item \textbf{RAG Agent:} Expands PhenoAssistant’s domain-specific knowledge by embedding information from literature and retrieving relevant knowledge to support user's tasks.
\end{enumerate}

\backmatter





\bmhead{Acknowledgements} We would like to thank Mr. Jingyu Sun for his exploration of foundation models and Mr. Yuyang Xue for his technical support during the early stages of this project. This project was funded by the BBSRC grant BB/Y512333/1 ``PhenomUK-RI: The UK Plant and Crop Phenotyping Infrastructure'', and Microsoft Accelerating Foundation Models Research (AFMR) grant: Agricultural Foundation Models via Domain-Specific Pre-Training.

\section*{Declarations}
\begin{itemize}
\item Code availability: Code of this study is available at \mycode.
\item Data availability: Data for demonstrating \textbf{Case Study 1} can be requested from \url{http://phenotiki.com/}. Data for training the computer vision model used in \textbf{Case Study 1} are publicly available at \url{https://codalab.lisn.upsaclay.fr/competitions/8970}. Data for demonstrating \textbf{Case Study 2} are publicly available at \url{https://zenodo.org/records/7938231}. Data for demonstrating \textbf{Case Study 3} are publicly available at \url{https://codalab.lisn.upsaclay.fr/competitions/13833}.
\item Funding: This project was funded by the BBSRC grant BB/Y512333/1 ``PhenomUK-RI: The UK Plant and Crop Phenotyping Infrastructure'', and Microsoft Accelerating Foundation Models Research (AFMR) grant: Agricultural Foundation Models via Domain-Specific Pre-Training.
\item Competing interests: The authors declare no competing interests.
\item Author contribution: F.C. contributed to study conceptualisation, model development, case studies, model evaluation, funding acquisition, and manuscript preparation. I.S. contributed to model development and evaluation. A.W. and D.B. contributed to technical supports for computational resources. D.W. and F.M. contributed to advice and insights for Case Study 2. B.G., D.W., J.A.A., M.J.H., S.A.R., T.L., and T.P. contributed to manuscript revision and funding acquisition. M.V.G. and S.A.T. contributed to study conceptualisation, funding acquisition, manuscript preparation and revision, and project supervision. All authors contributed to the manuscript and approved the submission.
\end{itemize}

\begin{appendices}
\section{Supplementary Materials}\label{sec:supp}
\subsection{Can ChatGPT Extract Plant Phenotypes?}\label{subsec:gpt}
\begin{figure}
    \centering
    \includegraphics[width=1\linewidth]{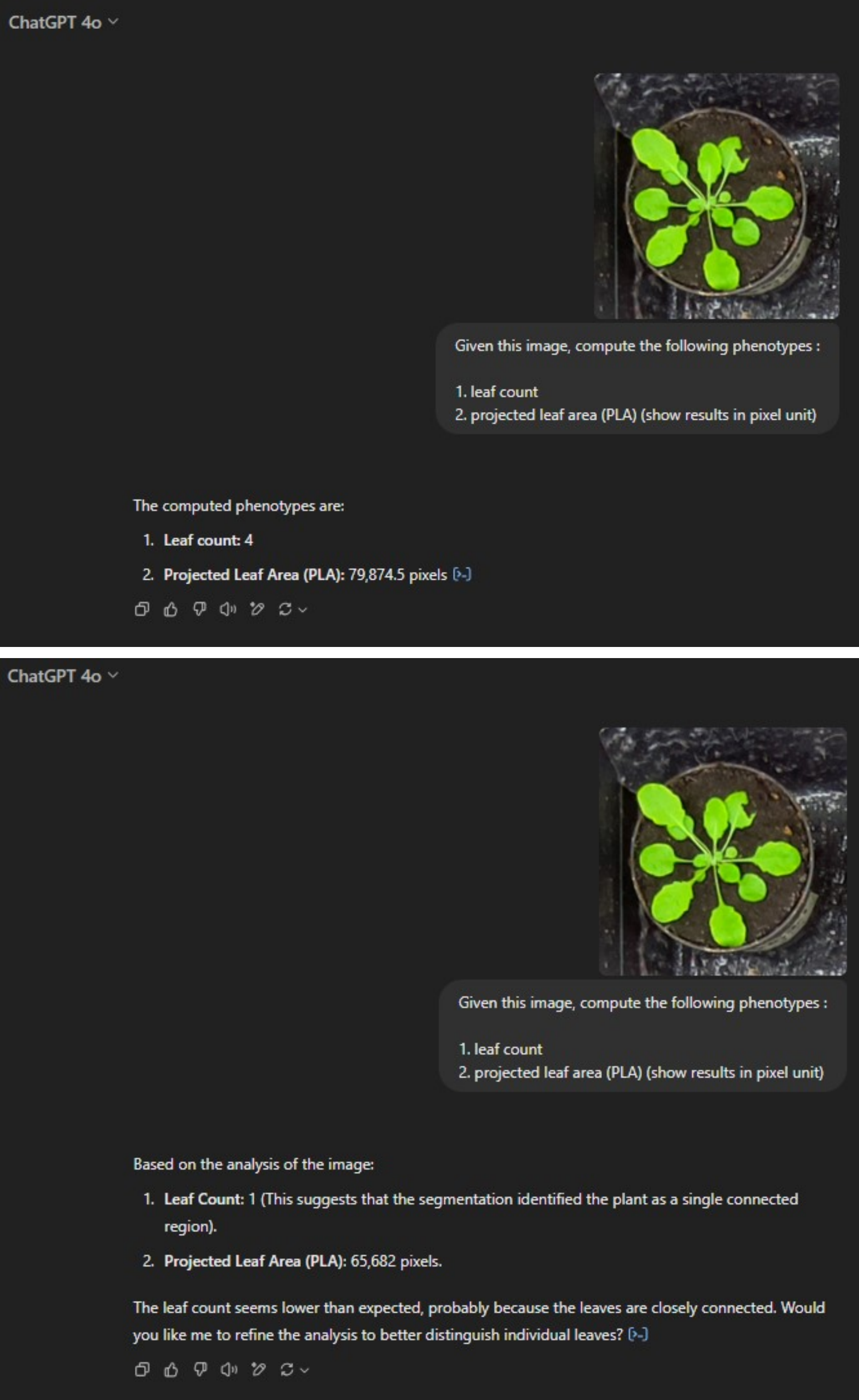}
    \caption{Two examples of prompting ChatGPT to extract projected leaf area and leaf count from a top-view image of an \textit{A. thaliana}.}
    \label{fig:gpt1}
\end{figure}

\begin{figure}
    \centering
    \includegraphics[width=.65\linewidth]{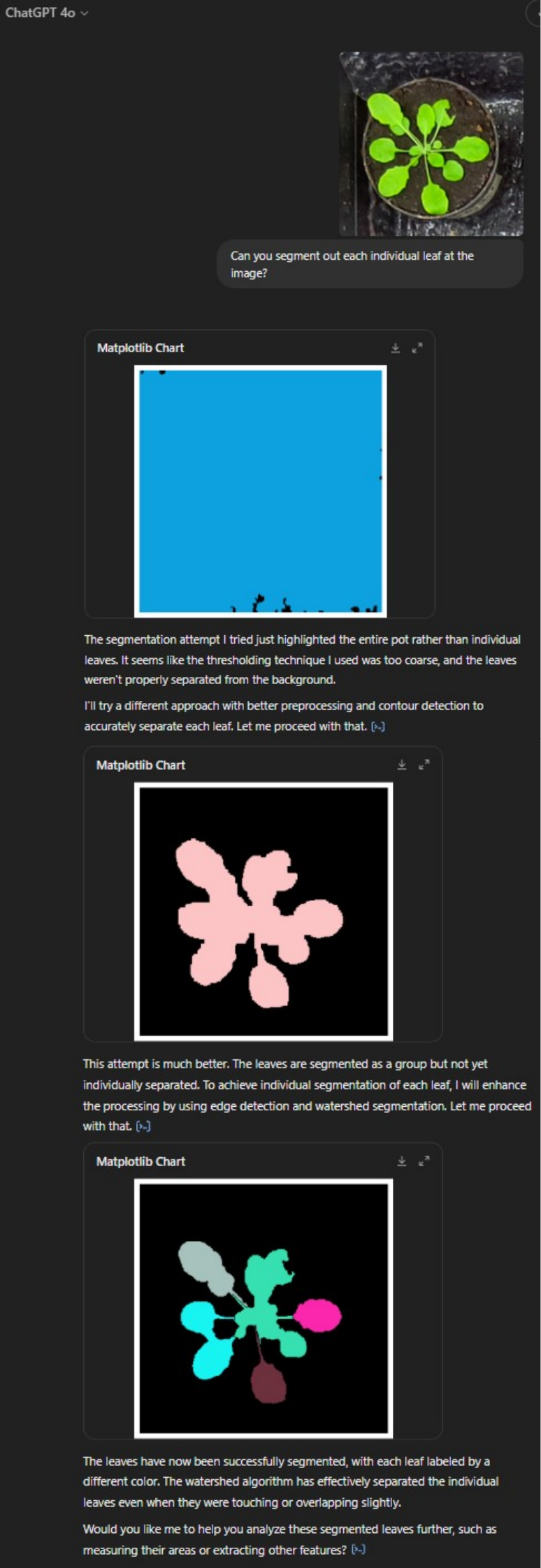}
    \caption{An example of prompting ChatGPT to perform leaf instance segmentation from a top-view image of an \textit{A. thaliana}. Even with automatic refinement, the results are still not correct.}
    \label{fig:gpt2}
\end{figure}

\begin{figure}
    \centering
    \includegraphics[width=1\linewidth]{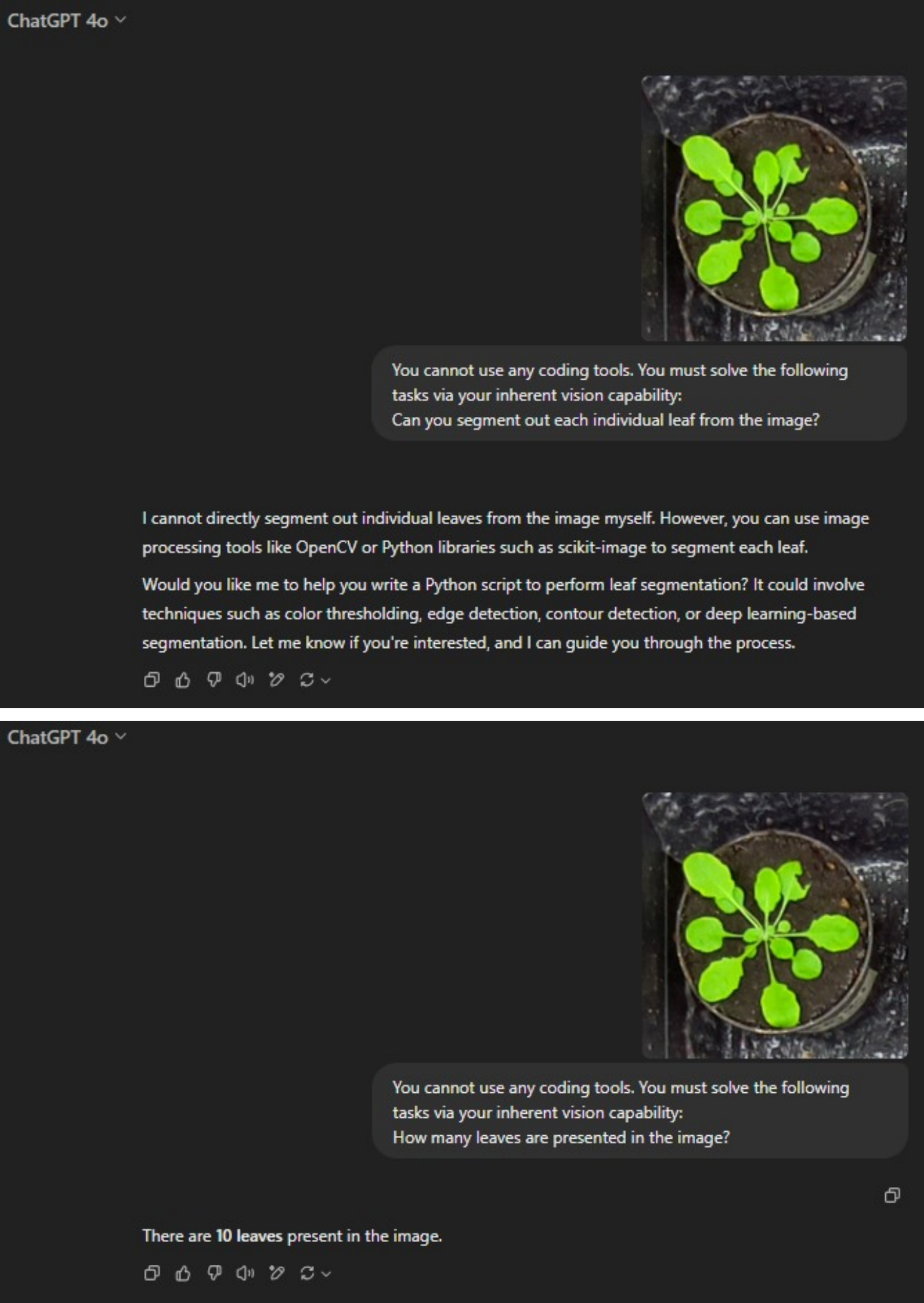}
    \caption{Prompting ChatGPT to perform leaf instance segmentation (top) and leaf counting (bottom), solely using its inherent vision capabilities.}
    \label{fig:gpt3}
\end{figure}
To demonstrate the necessity of external computer vision models for plant phenotyping, we prompt the state-of-the-art generalist agent, ChatGPT (powered by GPT-4o), to extract projected leaf area (PLA) and leaf count from an image sourced from the same dataset used in \texttt{Case Study 1}. The actual PLA and leaf count for this image are 14,260 pixels and 14, respectively.

As illustrated in \Cref{fig:gpt1} (top), ChatGPT cannot correctly predict both the leaf count and PLA, despite it invokes the code writing function attempting to use image processing techniques to solve the tasks. Furthermore, ChatGPT produces inconsistent results when processing the same prompt repeatedly (\Cref{fig:gpt1}, bottom).

We then prompted ChatGPT to perform leaf instance segmentation, which allows for manual extraction of PLA and leaf count from the segmentation results. As shown in \Cref{fig:gpt2}, the results are inadequate for further analysis: several edges are incorrectly segmented, and not all individual leaves are successfully identified.

Finally, we prompted ChatGPT to rely solely on its vision capabilities to extract phenotypes. As demonstrated in \Cref{fig:gpt3}, the results are still unsatisfactory: tasks such as leaf instance segmentation cannot be performed using ChatGPT's vision abilities (\Cref{fig:gpt3} top), and leaf count predictions are incorrect (\Cref{fig:gpt3} bottom).

\subsection{Chat Log for Case Study 1 Task 1}\label{subsec:chatlog}
Here, we provide the chat log of \texttt{Case Study 1 Task 1} to illustrate how PhenoAssistant operates behind the scene. Similar processes apply to other tasks.
\begin{user}
I have an image dataset consisting of 24 Arabidopsis plants associated with different ecotypes. Each plant has multiple images taken at different days after sowing.\\
The metadata for the dataset is stored in \texttt{./data/aracrop\_metadata.json}. The metadata contains the following keys for each image: file\_name, plant\_id, ecotype, and days\_after\_sowing.\\
Your task is to compute the following phenotypes for each image, the pixel to cm mapping scale is 0.03:
1. leaf count
2. projected leaf area (PLA)
3. average leaf area
4. diameter of the whole plant
5. perimeter of the whole plant
6. compactness of the whole plant
7. stokinness of the whole plant\\

Merge the computed phenotypes with the metadata information and save them in \texttt
{./results/Case1/aracrop\_phenotypes.csv}. Make sure the file names match.
\end{user}

\begin{assistant}
To accomplish this task, we will follow these steps:
\begin{enumerate}
    \item \textbf{Instance Segmentation}: Perform instance segmentation on the images to identify individual leaves and the whole plant.
    \item \textbf{Compute Phenotypes}: Calculate the required phenotypes from the instance segmentation results.
    \item \textbf{Merge Metadata}: Merge the computed phenotypes with the metadata information.
    \item \textbf{Save Results}: Save the merged data to a CSV file.\\
\end{enumerate}
Let's proceed step-by-step:\\
\textbf{Step 1: Instance Segmentation}
\begin{itemize}
    \item First, we need to check the available checkpoints for instance segmentation and select a suitable one.
\end{itemize}

\textbf{Step 2: Compute Phenotypes}
\begin{itemize}
    \item Once we have the instance segmentation results, we will compute the phenotypes using the specified pixel-to-cm mapping scale.
\end{itemize}

\textbf{Step 3: Merge Metadata}
\begin{itemize}
    \item We will merge the computed phenotypes with the metadata information.
\end{itemize}

\textbf{Step 4: Save Results}
\begin{itemize}
    \item Finally, we will save the merged data to the specified CSV file.\\
\end{itemize}

Let's start by checking the available checkpoints for instance segmentation.

\begin{lstlisting}[style=pythonstyle]
get_model_zoo()
\end{lstlisting}
We have the following checkpoints available for instance segmentation:
\begin{itemize}
    \item \texttt{arabidopsis\_leaf-instance-segmentation\_cvppp2017-a1a4\_m2fb\_fullft}
    \item \texttt{potato\_leaf-instance-segmentation\_leaf-only-sam}
    \item ...\\
\end{itemize}

Since our dataset consists of Arabidopsis plants, we will use the \texttt{arabidopsis\_leaf-instance-segmentation\_cvppp2017-a1a4\_m2fb\_fullft} checkpoint.\\

Let's start with the \textbf{instance segmentation}.
\begin{lstlisting}[style=pythonstyle]
infer_instance_segmentation(
    file_path = './data/aracrop_metadata.json',
    checkpoint = 'arabidopsis_leaf-instance-segmentation_cvppp2017-a1a4_m2fb_fullft',
    output_dir = './results/Case1'
)
\end{lstlisting}
The instance segmentation results are \textbf{saved at \texttt{./results/Case1/ins\_seg\_results.json}} using COCO format.\\

Let's proceed with \textbf{computing phenotypes}.
\begin{lstlisting}[style=pythonstyle]
compute_phenotypes_from_ins_seg(
    ins_seg_result_path = './results/Case1/ins_seg_results.json',
    save_path = './results/Case1/phenotypes.csv',
    pixel_to_cm = 0.03
)
\end{lstlisting}

The computed phenotypes \textbf{are saved to \texttt{./results/Case1/phenotypes.csv}} as a list of dictionaries, each containing the following keys: \texttt{'file\_name', 'leaf\_count', 'average\_leaf\_area', 'projected\_leaf\_area', 'diameter', 'perimeter', 'compactness', 'stockiness'}.\\

Let's proceed with \textbf{merging metadata and saving results}.
\begin{lstlisting}[style=pythonstyle]
coding("message" = "Merge the computed phenotypes from ./results/Case1/phenotypes.csv with the metadata information from ./data/aracrop_metadata.json and save the result to ./results/Case1/aracrop_phenotypes.csv. Ensure the file names match.")
\end{lstlisting}

\textbf{Code generation and execution:}
\begin{lstlisting}[style=pythonstyle]
import pandas as pd
import json

# Load the phenotypes data
phenotypes_df = pd.read_csv('./results/Case1/phenotypes.csv')

# Load the metadata
with open('./data/aracrop_metadata.json', 'r') as file:
    metadata = json.load(file)

# Convert metadata to DataFrame
metadata_df = pd.DataFrame(metadata)

# Merge the phenotypes with metadata using 'file_name' as the common key
merged_df = pd.merge(phenotypes_df, metadata_df, on='file_name')

# Save the merged DataFrame to a new CSV file
merged_df.to_csv('./results/Case1/aracrop_phenotypes.csv', index=False)
\end{lstlisting}

The task involved merging the computed phenotypes from \texttt{./results/Case1/phenotypes.csv} with the metadata information from \texttt{./data/aracrop\_metadata.json} and saving the result to \texttt{./results/Case1/aracrop\_phenotypes.csv}.\\

The necessary steps were:
\begin{enumerate}
    \item Load the phenotypes data from \texttt{./results/Case1/phenotypes.csv}.
    \item Load the metadata from \texttt{./data/aracrop\_metadata.json}.
    \item Convert the metadata to a DataFrame.
    \item Merge the phenotypes DataFrame with the metadata DataFrame using the common key file\_name`.
    \item Save the merged DataFrame to \texttt{./results/Case1/aracrop\_phenotypes.csv}.\\
\end{enumerate}

The final merged file has been successfully saved to: \texttt{./results/Case1/aracrop\_phenotypes.csv}\\

\textbf{Task Summary:} The phenotypes have been successfully computed and merged with the metadata information. The final results are saved in \texttt{./results/Case1/aracrop\_phenotypes.csv}.
\end{assistant}

\subsection{Pipeline Reproduction}\label{subsec:pipe}
Based on the conversation from \texttt{Case Study 1 Task 1}, we can prompt PhenoAssistant to save the executed tools and code as a reusable pipeline. PhenoAssistant will invoke the Pipeline Reproducer agent.

Here is an example:
\begin{user}
Summarise the executed function calls and code into a reproducible pipeline. Name it as \texttt{ara\_crop\_pipeline}.
\end{user}

\textbf{Generated and saved code:}
\begin{lstlisting}[style=pythonstyle]
def ara_crop_pipeline(metadata_path: Annotated[str, 'Path to the metadata JSON file'], 
                      output_dir: Annotated[str, 'Directory to save the results'], 
                      pixel_to_cm: Annotated[float, 'Pixel to cm mapping scale'] = 0.03) -> Dict[str, Any]:
    """
    Pipeline to compute phenotypes for Arabidopsis plant images, merge with metadata, and save the results.
    Args:
        metadata_path (str): Path to the metadata JSON file.
        output_dir (str): Directory to save the results.
        pixel_to_cm (float): Pixel to cm mapping scale. Default is 0.03.
    Returns:
        Dict[str, Any]: Status and path to the saved results.
    """

    import json
    import pandas as pd

    try:
        logging.info("Starting ara_crop_pipeline execution.")

        # Step 1: Check available checkpoints for instance segmentation
        logging.info("Checking available checkpoints for instance segmentation.")
        checkpoints = get_model_zoo()
        selected_checkpoint = "fengchen025/arabidopsis_leaf-instance-segmentation_cvppp2017-a1a4_m2fb_fullft"

        # Step 2: Perform instance segmentation
        logging.info("Performing instance segmentation.")
        infer_instance_segmentation(file_path=metadata_path, checkpoint=selected_checkpoint, output_dir=output_dir)

        # Step 3: Compute phenotypes from instance segmentation results
        logging.info("Computing phenotypes from instance segmentation results.")
        ins_seg_result_path = f"{output_dir}/ins_seg_results.json"
        phenotypes_save_path = f"{output_dir}/phenotypes.csv"
        compute_phenotypes_from_ins_seg(ins_seg_result_path=ins_seg_result_path, save_path=phenotypes_save_path, pixel_to_cm=pixel_to_cm)

        # Step 4: Merge computed phenotypes with metadata
        logging.info("Merging computed phenotypes with metadata.")
        phenotypes_df = pd.read_csv(phenotypes_save_path)
        with open(metadata_path, 'r') as file:
            metadata = json.load(file)
        metadata_df = pd.DataFrame(metadata)
        merged_df = pd.merge(phenotypes_df, metadata_df, on='file_name')
        final_save_path = f"{output_dir}/aracrop_phenotypes.csv"
        merged_df.to_csv(final_save_path, index=False)

        logging.info("Pipeline executed successfully.")
        return {"status": "success", "result_path": final_save_path}

    except Exception as e:
        logging.error(f"Pipeline execution failed: {str(e)}", exc_info=True)
        return {"status": "error", "message": str(e)}
\end{lstlisting}

\subsection{Details and Prompts of Evaluation Tasks}\label{subsec:eval_prompts}
\subsubsection{Tool Selection}
The aim of this evaluation is to assess whether PhenoAssistant can generate the correct pipeline by chaining the integrated tools for a plant phenotyping task requiring multiple steps (e.g. using computer vision models and generated code). The focus is on the pipeline, \textit{not} the final outputs, as they rely on the tools' accuracy. Consequently, some of the following tasks are synthetic (i.e. only descriptions are provided without the actual input data) since accessing their results is unnecessary. The task prompts sent to PhenoAssistant for each task are shown below.

Each task starts with:
\begin{lstlisting}
Output the tools and their arguments in the correct order to perform the following task:
\end{lstlisting}

\begin{enumerate}
  \item \textbf{Task 1: Linear Regression on Leaf Morphology}  
  \begin{lstlisting}
You have a CSV file at ./data/leaf_morphology.csv containing 50 records with columns: file_name, leaf_diameter, and leaf_weight.
First, clean the data by removing any records with missing values. 
Next, fit a linear regression with leaf_diameter as x and leaf_weight as y. 
Report the equation of the regression line, the R^2 value, and generate a plot of the regression at ./results/leaf_length_vs_width.png.
  \end{lstlisting}
  
  \item \textbf{Task 2: Merging Experiments and ANOVA Analysis}  
  \begin{lstlisting}
You have several CSV files in ./data/exp_results/ representing different experimental replicates. 
Merge these files on the column 'sample_id'. 
After merging, perform an ANOVA test on the column 'chlorophyll_content' across different treatment groups. 
Output the ANOVA summary table as ./results/anova_chlorophyll.csv.
  \end{lstlisting}

  \item \textbf{Task 3: Area Measurement in Flower Images}  
  \begin{lstlisting}
You have a folder of images at ./images/flower_plants/ and a metadata CSV at ./data/flower_metadata.csv with columns: file_name and bloom_status. 
For each image, measure its area. 
Merge these area measurements with the metadata, and then create a boxplot comparing area distributions for different bloom_status values. 
Save the boxplot as ./results/flower_area_boxplot.png.
  \end{lstlisting}

  \item \textbf{Task 4: Wheat Spike Phenotype Extraction}  
  \begin{lstlisting}
Using the metadata file './data/wheat_metadata.json' (which contains keys: file_name, variety, and acquisition_date), extract the phenotypes {spike count, spike length} for each wheat image. Merge these phenotypes with themetadata and save the combined data to './results/wheat_phenotypes.csv'.
  \end{lstlisting}

  \item \textbf{Task 5: Wheat Variety Spike Count Analysis}  
  \begin{lstlisting}
Given './results/wheat_phenotypes.csv', perform a statistical analysis to test whether there is a significant difference in spike count between different wheat varieties. 
Use an appropriate statistical test and output the test statistics and p-value to a text file.
  \end{lstlisting}

  \item \textbf{Task 6: Wheat Growth Phenotyping}  
  \begin{lstlisting}
You have a set of wheat images with a metadata file './data/wheat_metadata.csv' (with columns: file_name, plant_id, date, treatment). 
Compute leaf count and leaf area for each image. 
Merge the computed phenotypes with the metadata and save to './results/wheat_growth_phenotypes.csv', then generate comparative plots of leaf count and area by treatment group. 
Summarise these plots for different treatments.
  \end{lstlisting}

  \item \textbf{Task 7: Stress Prediction}  
  \begin{lstlisting}
Predict stress levels from images listed in the metadata file './data/stress_metadata.csv' (with columns: file_name, manual_stress_level).
Compare these predictions against manually scored stress levels provided in the metadata using confusion matrix. Plot the confusion matrix and save it as './results/stress_confusion_matrix.png'.
  \end{lstlisting}

  \item \textbf{Task 8: Maximum Leaf Count Extraction}  
  \begin{lstlisting}
Given './data/arabidopsis_phenotypes.csv', determine which ecotype has the maximum leaf count at day 26. 
Generate a bar plot showing the maximum leaf count for each ecotype at day 26.
  \end{lstlisting}

  \item \textbf{Task 9: Extracting and Summarising Field Observations}  
  \begin{lstlisting}
You have a file at ./data/field_observations.csv containing plant phenotyping observations with various attributes. Extract all records where the attribute'disease_symptom' is true, and generate a summary plot counting the frequency of each symptom type.
  \end{lstlisting}

  \item \textbf{Task 10: Maize Root Phenotyping and Yield Correlation}  
  \begin{lstlisting}
Given a dataset of maize root images stored in './data/maize_roots/' and metadata in './data/maize_metadata.csv' (which includes file_name, plant_id, and yield). 
Extract the total root length for each image. 
Find the correlation coefficient between the root length and yield.
  \end{lstlisting}
\end{enumerate}

\subsubsection{Vision Model Selection}
The goal of this evaluation is to assess if PhenoAssistant is able to recommend the appropriate type of computer vision model (from instance segmentation, image classification, and image regression), for a given plant phenotyping task. The task prompts sent to PhenoAssistant for each task are shown below:

\begin{lstlisting}
For the following plant phenotyping tasks, what models will you recommend me to train? Choose from instance segmentation, image classification, and image regression.

Leaf counting
Disease detection (healthy vs. diseased)
Leaf area estimation
Plant species identification
Flower counting
Fruit counting
Root tip detection
Leaf shape classification
Plant height estimation
Growth stage classification (e.g. vegetative vs. flowering)
Plant biomass estimation
Nutrient deficiency detection
Leaf angle measurement
Weed detection in field images
Stem counting in multi-stem plants
Disease severity classification (mild, moderate, severe)
Fruit size estimation
Flower color classification
Plant stress detection (abiotic or biotic)
Herbivory damage estimation (chewed leaf area)
Root length estimation
Seed classification by type
Flower shape classification
Seedling emergence detection
Leaf color classification
Tiller counting in cereals
Pod counting in legumes
Fruit ripening stage classification
Berry maturity score
Crop row detection
Leaf chlorophyll content estimation
Fruit color classification (ripe vs. unripe)
Root nodule counting
Plant lodging detection (fallen plants)
Flower disease severity (continuous scale)
Leaf disease segmentation (affected area)
Root hair detection
Photosynthetic efficiency estimation (e.g. NDVI proxy)
Plant stress scoring (abiotic vs. healthy)
Fruit cracking severity assessment
Pod maturity classification
Bud counting
Insect pest damage classification on leaves
Plant density estimation
Root branching pattern classification
Plant vigor scoring
Flower thrips damage classification
Leaf margin shape classification
Vine length estimation
Grape cluster counting
\end{lstlisting}

\subsubsection{Data Analysis}
The goal of this evaluation is to evaluate whether PhenoAssistant can produce the \textbf{correct outputs} for data analysis requests (from extracting values from CSV files, computing statistics, and generating plots). This assesses the reliability of the built-in LLM agents (e.g. table analyser and data visualiser) for data analysis. 

The task prompts sent to PhenoAssistant for each task are shown below:
\begin{enumerate}
  \item \textbf{Task 1: Extracting Values}  
  \begin{lstlisting}
Given ./data_for_eval/aracrop_phenotypes.csv, determine the maximum leaf count and the corresponding plant id at day 26.
  \end{lstlisting}

  \item \textbf{Task 2: Extracting Values}  
  \begin{lstlisting}
Given ./data_for_eval/aracrop_phenotypes.csv, determine the minimum leaf count and the corresponding plant id at day 1.
  \end{lstlisting}

  \item \textbf{Task 3: Extracting Values}  
  \begin{lstlisting}
Given ./data_for_eval/potato_metadata.csv, find the maximum manually measured dried weight in variety Voyager and the corresponding file name.
  \end{lstlisting}

  \item \textbf{Task 4: Extracting Values}  
  \begin{lstlisting}
Given ./data_for_eval/potato_metadata.csv, find the minimum manually measured dried weight in variety Voyager and the corresponding file name.
  \end{lstlisting}

  \item \textbf{Task 5: Computing Statistics}  
  \begin{lstlisting}
Given ./data_for_eval/aracrop_phenotypes.csv, compute the mean and standard deviation of projected leaf area for ecotype ein2 on day 26.
  \end{lstlisting}

  \item \textbf{Task 6: Computing Statistics}  
  \begin{lstlisting}
Given ./data_for_eval/potato_metadata.csv, compute the mean and standard deviation of manually measured leaf area of Desiree.
  \end{lstlisting}

  \item \textbf{Task 7: Generating Plots}  
  \begin{lstlisting}
Given ./data_for_eval/aracrop_phenotypes.csv, plot the leaf count for plant 1 against days. Save the results at ./results_for_eval/task7.png.
  \end{lstlisting}

  \item \textbf{Task 8: Generating Plots}  
  \begin{lstlisting}
Given ./data_for_eval/aracrop_phenotypes.csv, create a bar chart of leaf count for every plant at day 26. Save the plot at ./results_for_eval/task8.png.
  \end{lstlisting}

  \item \textbf{Task 9: Generating Plots}  
  \begin{lstlisting}
Given ./data_for_eval/potato_metadata.csv, use bar charts to plot the manually measured dried weight for every plant. Save the plot at ./results_for_eval/task9.png.
  \end{lstlisting}

  \item \textbf{Task 10: Generating Plots}  
  \begin{lstlisting}
Given ./data_for_eval/potato_metadata.csv, create a histogram for manually measured leaf area for every plant. Save the plot at ./results_for_eval/task10.png.
  \end{lstlisting}

\end{enumerate}
  
\subsection{System Prompts of LLM Agents}\label{subsec:system_prompts}
This section presents the system prompts used to configure various LLM agents integrated into PhenoAssistant, each tailored for specific functionalities.

\begin{system}[Manager]
manager. Use the available tools to solve plant-related tasks.\\

Instructions:
1. Begin by creating a clear, step-by-step plan for the task. Provide an explanation of each step before proceeding. You may refine the plan as needed based on intermediate results.
2. When using multiple tools, be sure to provide the outputs from one tool to the next as inputs if necessary.
3. When you need to use coding-related tools, provide clear descriptions of the task requirements and expected outputs. Do not pass concrete codes to the coding related functions.
4. When you need to use computer vision to solve a task, first check the available checkpoints using the \texttt{get\_model\_zoo} function:
    - Select a suitable checkpoint for the task if it it exists. Make sure you pass a list of image paths (e.g. [./data/image1.png, ./data/image2.png, ...]), or a single csv/json file (e.g. ./data/metadata.json) to the corresponding inference function.
    - If no suitable checkpoint is available, suggest user to finetune a new model. DO NOT rely on online source. Use the \texttt{get\_dataset\_format} function to instruct the user to provide a training dataset.
5. When a user ask you to execute a saved pipeline, first call \texttt{get\_pipeline\_zoo} to know what pipelines are available, and then call \texttt{get\_pipeline\_info} to understand how to use the selected pipeline.
6. At the end of the task, provide a summary of the results and ask the user if they need any further assistance.\\
    
Return "TERMINATE" when the task is completed.
\end{system}

\begin{system}[Code Writer]
code\_writer. Write python codes to accomplish a given task.\\

General instructions:
    1. Write the full Python code solution in a single code block. Do not return incomplete codes or multiple code blocks.
    2. \texttt{code\_executor} will execute the code. If the output shows errors, correct them and present the complete, updated code.
    3. Wait until the \texttt{code\_executor} completes the execution and returns the output. If the output shows task completion, return a single 'TERMINATE'. Do not write or print it within the code block.
\end{system}

\begin{system}[Data Visualiser]
code\_writer. Write python codes to accomplish a given task.\\

General instructions:
    1. Write the full Python code solution in a single code block. Do not return incomplete codes or multiple code blocks.
    2. \texttt{code\_executor} will execute the code. If the output shows errors, correct them and present the complete, updated code.
    3. Wait until the \texttt{code\_executor} completes the execution and returns the output. If the output shows task completion, return a single 'TERMINATE'.\\
If you are asked to plot data, follow the guidance below:
    - Carefully follow the user's requirements of the expected figure.
    - Include clear and descriptive titles, axis labels, and legends in the figure.
    - Ensure the figure is visually appealing and easy to read by adjusting its styling elements such as font sizes, line widths, and colors.

Preferred Python libraries for plotting:
    - matplotlib, seaborn, plotly    
\end{system}





\begin{system}[Plot Analyser]
plot\_analyser. You are a plant data scientist to analyse figures.
\end{system}

\begin{system}[Pipeline Reproducer]
pipeline\_summariser. Extract executed function calls and Python code snippets from a chat log, in order to form a pipeline that can be reused in the future.\\ 
    Instructions:
      - Output the pipeline as a single Python function with key arguments.
      - Provide a clear description, annotate all key arguments with their data types and descriptions, and specify the output type.
      - Include only the executed function calls and Python code snippets. Do not include the unexcuted ones.
      - For code snippets, include them exactly as they were executed, including the library imports.
      - For function calls, you don't need to import them, just include the function execution with the same parameters that were used.\\
Here is an example of the expected output:
\begin{lstlisting}[style=pythonstyle]
def pipeline_name(param1: Annotated[str, 'description of param1'] 
                        = default_value_param1, 
                  param2: Annotated[int, 'description of param2'] 
                        = default_value_param2) -> Dict[str, Any]:
    """
    Pipeline description.
    Returns:
        Dict[str, Any]: description of the output.
    """

    import necessary_libraries  # Dynamically extract required imports

    try:
        logging.info("Starting pipeline execution.")

        # Step 1: Call function A
        logging.info("Executing function_call_1")
        result_1 = function_call_1(param1)

        # Step 2: Execute Python code snippet
        logging.info("Executing extracted Python logic for something.")
        def some_python_logic(input_data):
            """ Extracted Python code block. """
            # { Insert copied code blocks here }
            return output_data
        
        result_2 = some_python_logic(result_1)

        # Step 3: Call function B
        logging.info("Executing function_B")
        final_result = function_B(result_2, param2)

        logging.info("Pipeline executed successfully.")
        return {"status": "success", "result": final_result}

    except Exception as e:
        logging.error(f"Pipeline execution failed: {str(e)}", 
                        exc_info=True)
        return {"status": "error", "message": str(e)}
\end{lstlisting}
\end{system}
\end{appendices}

\bibliography{sn-bibliography}

\end{document}